\documentclass[letterpaper,english,prl,twocolumn]{revtex4}
\usepackage[T1]{fontenc}
\usepackage[latin1]{inputenc}
\usepackage{subfigure}
\usepackage{graphicx}

\makeatletter

\usepackage{babel}
\makeatother
\begin{document}

\preprint{This line only printed with preprint option}

\title{Prediction of Superconductivity at $\approx30$ K in Compressed Body-Centered
Cubic Yttrium}

\author{Prabhakar P. Singh}

\email{ppsingh@phy.iitb.ac.in}

\affiliation{Department of Physics, Indian Institute of Technology, Powai, Mumbai-
400076, India }

\begin{abstract}
Using \emph{ab initio} methods, we have studied the electron-phonon
interaction in compressed, body-centered cubic (bcc) yttrium, which
is predicted to be stable at $280$ GPa {[}Melsen \emph{et al}, Phys.
Rev. B \textbf{48}, 15574 (1993)]. We find that compressed, bcc yttrium
has a large electron-phonon coupling with $\lambda=1.8$, leading
to a superconducting transition temperature $T_{c}\approx30$ K or
above. Our results indicate that the large electron-phonon coupling
is due to the lattice hardening. 
\end{abstract}
\maketitle
Recent experiments on pressure-induced increase in superconducting
transition temperature $T_{c}$ of yttrium \cite{hamlin} seem to
suggest an almost continuous increase in $T_{c}$ with pressure. With
increase in pressure from ambient to $115$ GPa, the structural changes
\cite{gro_dhcp,vohra} in yttrium from hcp$\rightarrow$Sm-type$\rightarrow$dhcp$\rightarrow$trigonal
are accompanied by increase in $T_{c}$ from $6$ mK \cite{probst}
to $20$ K \cite{hamlin}. In contrast, $T_{c}$ increases with pressure
only in the fcc phase of lithium as it changes from bcc$\rightarrow$fcc$\rightarrow$hR1$\rightarrow$cI16
with pressure going up to $70$ GPa \cite{hanfland}. Thus, the pressure-induced
changes in $T_{c}$ seem to be intimately connected with the structural
phase transitions in these metals \cite{asch}. 

It has been shown theoretically that yttrium would stabilize in a
bcc phase above $280$ GPa \cite{melsen}. Taking a cue from the experiments
on yttrium \cite{hamlin}, it is not surprising to expect bcc yttrium
to superconduct well above $20$ K at pressures above $280$ GPa.
However, it is also possible that in this pressure range yttrium may
start to behave like lithium \cite{shimuzu,stru,hanfland,profeta,deepa}.

To further investigate the changes in the superconducting properties
of bcc yttrium, we have used density-functional-based methods to calculate
its electronic structure, lattice-dynamical response and the electron-phonon
interaction. Subsequently, we have solved the isotropic Eliashberg
gap equation to obtain the superconducting transition temperature
$T_{c}$. 

Based on our calculations, we find that compressed, bcc yttrium has
a large electron-phonon coupling with $\lambda=1.8$, leading to a
superconducting transition temperature $T_{c}\approx36$ K for $\mu^{*}=0.25$.
Such a large coupling is a direct consequence of lattice hardening
and its subsequent coupling to Fermi electrons. We also find the electrons
at Fermi energy to have predominantly $t_{2g}$ symmetry. Before we
describe our results, we briefly outline the computational details
of our calculations. Further details can be found in our previous
work \cite{pps_y1} as well as the references therein. The notations
used here are that of our previous work \cite{pps_y1}.

We have calculated the electronic structure, the phonon density of
states, $F(\omega)$, and the Eliashberg function, $\alpha^{2}F(\omega)$
of bcc yttrium using full-potential, linear muffin-tin orbital (LMTO)
method \cite{savrasov1,savrasov2}. Subsequently, we have numerically
solved the isotropic Eliashberg gap equation \cite{allen1,allen2,private1}
for a range of $\mu^{*}$ to obtain the corresponding $T_{c}.$ 

The lattice constant $a=5.15$ \emph{a.u.} used in our calculations
for bcc yttrium approximately corresponds to a compressed volume equal
to $0.31v_{0}$, where $v_{0}$ is the experimentally determined equilibrium
volume of hcp yttrium \cite{gro_dhcp,vohra,pps_y1}.

The charge self-consistent, full-potential, LMTO calculations for
bcc yttrium was carried out with $2\kappa-$energy panel, 897 $\mathbf{k}$-points
in the irreducible wedge of the Brillouin zone, and the muffin-tin
radius for yttrium was taken to be 2.22 atomic units. The 4\emph{s}
state of yttrium was treated as a semicore state, while the 4\emph{p}
state was treated as a valence state. The basis set used consisted
of \emph{s}, \emph{p}, and \emph{d} orbitals at the yttrium site,
and the potential and the wave function were expanded up to $l_{max}=6$. 

The Fermi surface, phonon density of states, and the Eliashberg function
of bcc yttrium were calculated using the  linear response code based
on the full-potential, LMTO method. The Fermi surface was constructed
using XCrySDen program \cite{xcrys} with eigenvalues calculated on
a $48\times48\times48$ grid in the reciprocal space. The dynamical
matrices and the electron-phonon matrix elements were calculated on
a $8\times8\times8$ grid resulting in $29$ irreducible $\mathbf{q}$-points.
The Brillouin zone integrations during linear response calculations
were carried out using a $16\times16\times16$ grid of $\mathbf{k}$-points.
The Fermi surface sampling for the evaluation of the electron-phonon
matrix elements was done using a $48\times48\times48$ grid. Another
set of calculations of self-consistent charge density, $F(\omega)$
and $\alpha^{2}F(\omega)$ was carried out at a lattice constant $a=5.1987$
a.u. using a $6\times6\times6$ grid with $16$ irreducible $\mathbf{q}$-points.
In this case, a $12\times12\times12$ grid of $\mathbf{k}$-points
was used for linear-response calculations, in conjunction with a $36\times36\times36$
grid for Fermi surface sampling. The muffin-tin radius for yttrium
was taken to be 2.251 atomic units. For visualizing the charge density
within a $5$ mRy energy window to the Fermi energy we used the atomic-sphere-approximation
variant of the LMTO code \cite{tblmto}. 

\begin{figure}
\begin{centering}\subfigure[]{\includegraphics[clip,scale=0.3]{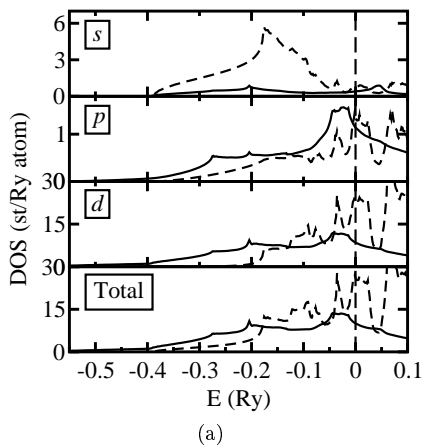}}\par\end{centering}

\caption{The \emph{s}, \emph{p}, \emph{d} and the total densities of states
of bcc (solid line) and hcp (dashed line) yttrium. The vertical dashed
line indicates the Fermi energy. }
\end{figure}

In Fig. 1, we show the  total and the \emph{l}-resolved densities
of states (DOS) of bcc yttrium. For comparison, we have also included
the DOS for the hcp yttrium \cite{pps_y1}. From Fig. 1, we see that
the pressure-induced pushing out of \emph{s}-electrons is followed
by creation of new \emph{d}-states between $-0.55$ Ry and $-0.25$
Ry to accommodate the \emph{s}-electrons. The flattening of the \emph{d}-band
reduces the DOS at the Fermi energy to $10.03$ st/Ry from its hcp
value of $25.89$ st/Ry, with most of the contributions coming from
the \emph{d}-states. 

\begin{figure}
\begin{centering}\includegraphics[clip,scale=0.3]{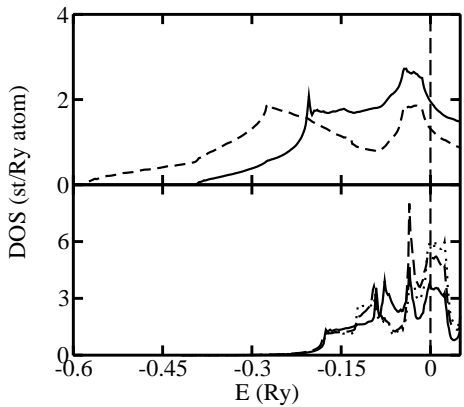}\par\end{centering}

\caption{The symmetry-resolved \emph{d} densities of states in bcc (top panel)
and hcp (bottom panel) yttrium. The densities of states in the top
panel correspond to $t_{2g}$ ($xy$, y$z$ and $zx$; solid line)
and $e_{g}$ ($x^{2}-y^{2}$and $3z^{2}-1$; dashed line) symmetries,
while in the bottom panel they correspond to $yz$ ($zx$; solid line),
$xy$ ($x^{2}-y^{2}$; dashed line) and $3z^{2}-1$ (dotted line)
symmetries. The vertical dashed line indicates the Fermi energy. }
\end{figure}

It is interesting to consider the symmetry-resolved \emph{d}-DOS,
as shown in Fig. 2. Some of the differences in the DOS between the
ambient pressure hcp phase and the high-pressure bcc phase arise due
to change in symmetry from hexagonal to cubic. In hcp phase, the three
independent \emph{d}-DOS corresponding to $yz$ ($zx$), $xy$ ($x^{2}-y^{2}$)
and $3z^{2}-1$ symmetries have similar bandwidths but some differences
in the DOS as a function of energy. However, in bcc phase the bottom
of the doubly degenerate $e_{g}$ ($x^{2}-y^{2}$ and $3z^{2}-1$)
symmetry DOS is almost $0.2$ Ry below the bottom of the triply degenerate
$t_{2g}$ ($xy$, y$z$ and $zx$) symmetry DOS, indicating transfer
of pushed-out \emph{s}-electrons to $e_{g}$ symmetry orbitals. A
symmetry-resolved breakup of DOS at the Fermi energy is $0.40$, $1.14$,
$5.87$ and $2.62$ st/Ry for $s$, $p$, $t_{2g}$ and $e_{g}$ symmetries,
respectively. 

\begin{figure}
\begin{centering}\includegraphics[clip,scale=0.3,angle=270]{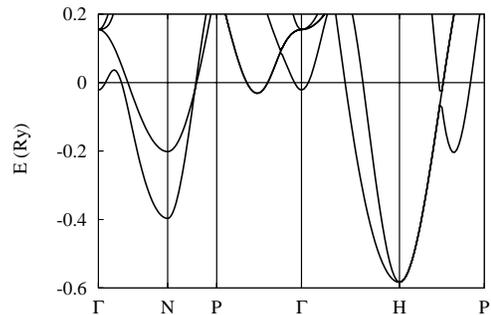}\par\end{centering}

\begin{centering}\par\end{centering}

\caption{The band-structure of bcc yttrium (solid line) along high symmetry
directions in the corresponding Brillouin zone. The horizontal line,
passing through the energy zero, indicates the Fermi energy. }
\end{figure}

As a prelude to understanding the Fermi surface topology and the nature
of electrons contributing to the electron-phonon interaction, we show
in Fig. 3, the calculated band-structure of bcc yttrium along high
symmetry directions in the corresponding Brillouin zone. The band
corresponding to 4\emph{p} state is much lower in energy, and hence
it is not shown in the figure. Based on Fig. 3 and an analysis of
the eigenvectors (not shown in the figure), the so-called fat bands,
we find that the bands crossing the Fermi energy along $\Gamma$-N,
N$-$P and H$-$P symmetry directions have substantial $t_{2g}$ symmetry
states, indicating an important role for these electrons as far as
the electron-phonon interaction was concerned.

Since Fermi surface plays an important role in understanding many
of the superconducting properties of materials, we show the calculated
Fermi surface of bcc yttrium in Fig. 4. The Fermi surface due to band
number 1 surrounds $\Gamma$ point and it has a bucket-like structure
around P symmetry point in the reciprocal space. The second band has
an umbrella-like structure with four-fold symmetry around N point,
in addition to having surfaces around $\Gamma$ and P points. It turns
out that some of the flat portions in all these surfaces lead to Fermi
surface nesting especially along N$-$P and $\Gamma$$-$P symmetry
directions, the indications of which can be seen in Figs. 4 (c) and
(d), showing the Fermi surface cut in a plane parallel to (110) plane.
A more detailed analysis is needed to clearly identify all the nesting
vectors.

\begin{figure}
\begin{centering}\subfigure[]{\includegraphics[clip,scale=0.22]{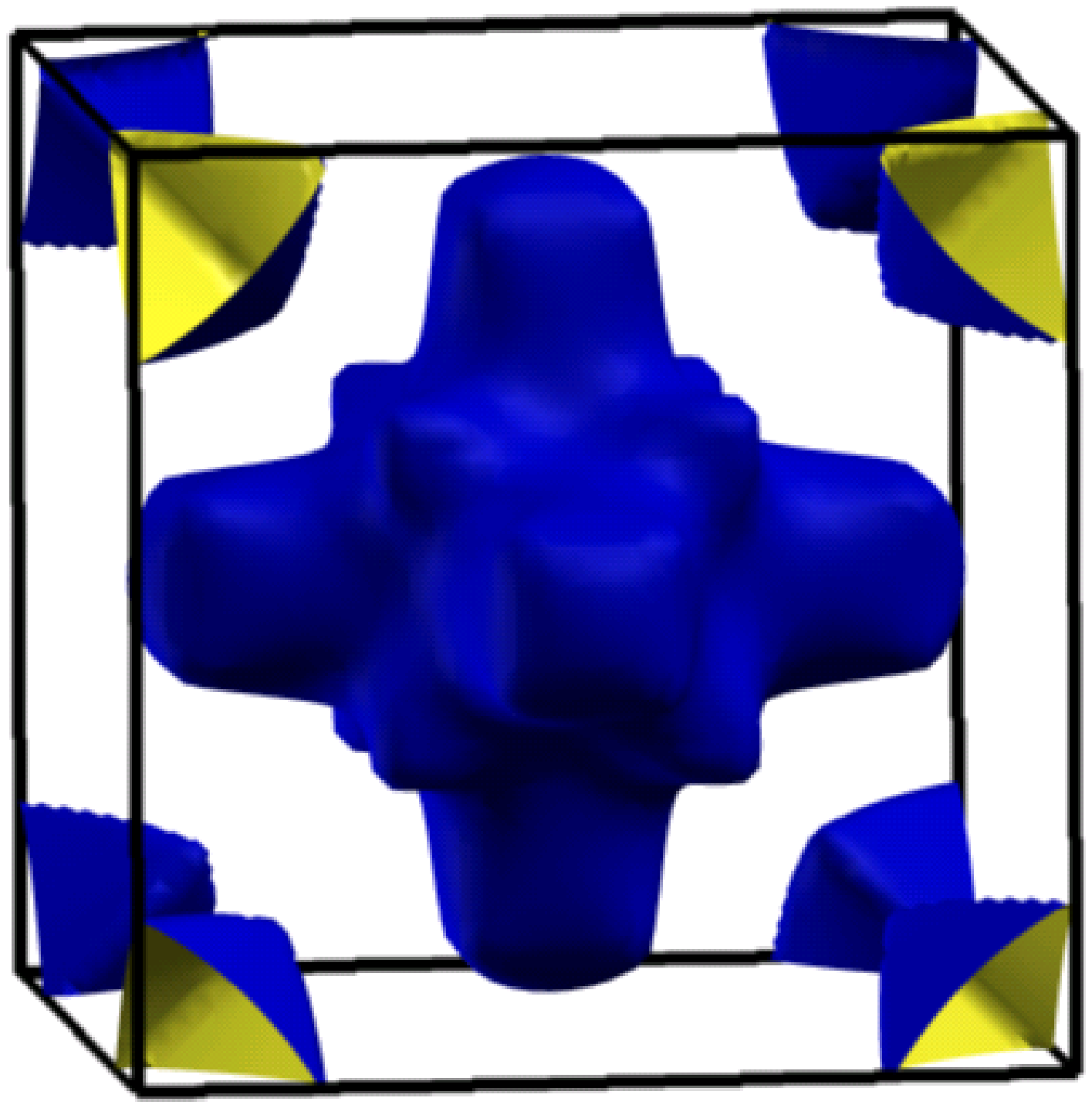}}
~~~\subfigure[]{\includegraphics[scale=0.25]{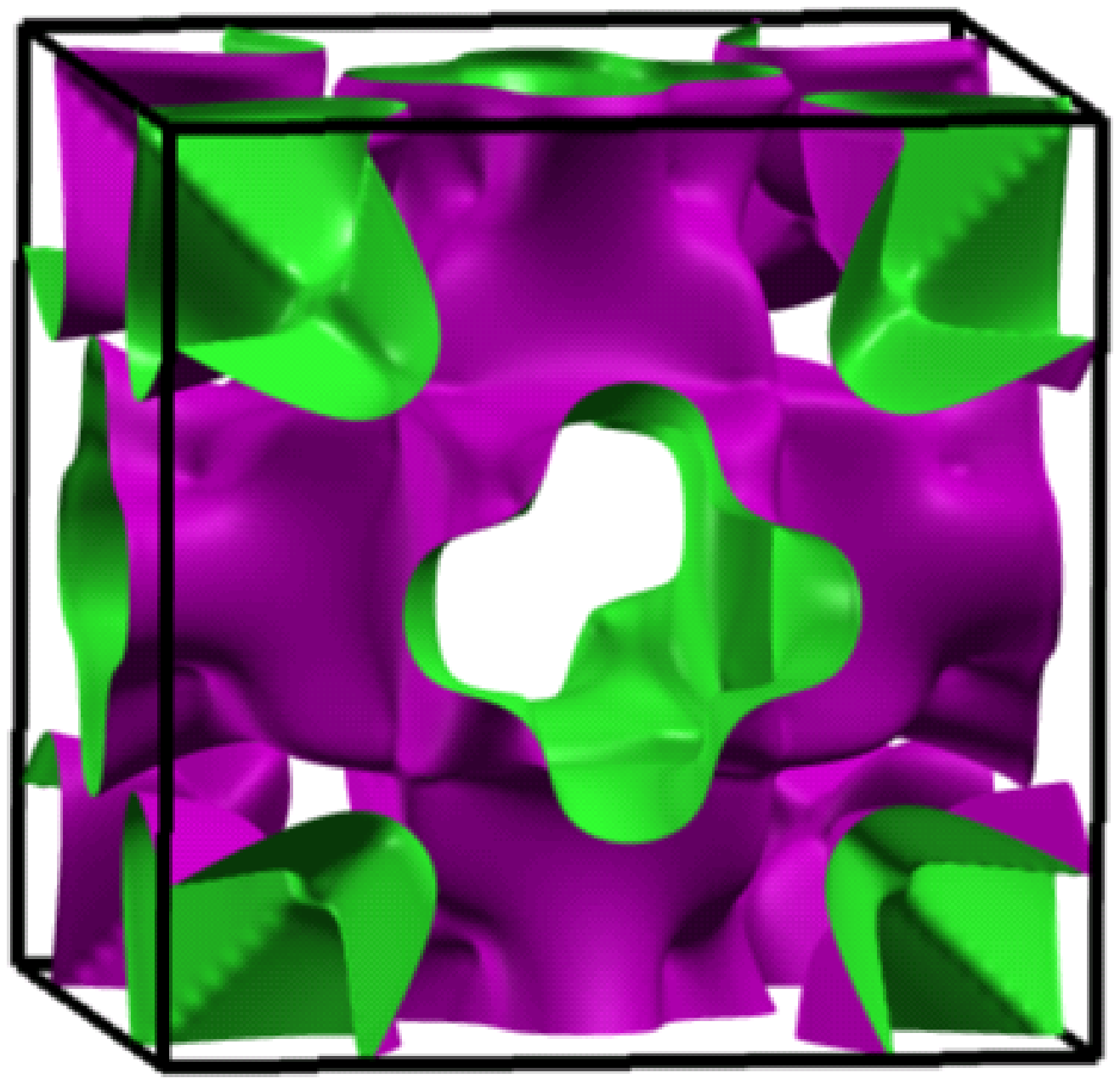}}\par\end{centering}

~

\begin{centering}\subfigure[]{\includegraphics[scale=0.5]{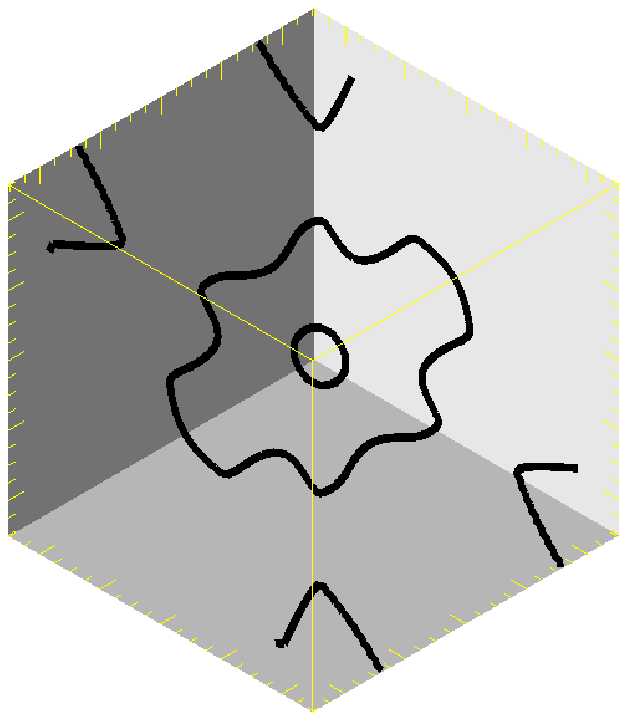}}~~~\subfigure[]{\includegraphics[scale=0.5]{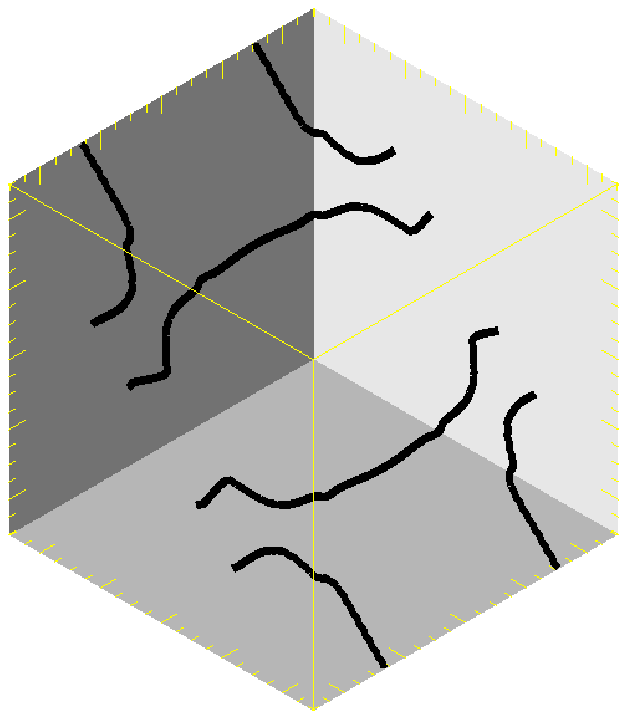}}\par\end{centering}

\caption{The Fermi surface of bcc yttrium in reciprocal space unit cell corresponding
to (a) band number 1, and (b) band number 2. The Fermi surface cut
on a plane parallel to (110) plane for (c) band number 1, and (d)
band number 2. The center of the cube corresponds to the $\Gamma$
point.}
\end{figure}

In Fig. 5 we show the charge density within a 5 mRy energy window
around the Fermi energy in bcc yttrium, calculated using the tight-binding-LMTO
method as mentioned earlier \cite{tblmto}. The $e_{g}$ symmetry
electrons seem to be localized around the yttrium sites while $t_{2g}$
electrons are more spread out and hybridized with the nearby yttrium
atoms. Having described the electronic structure of bcc yttrium, we
turn to its lattice-dynamical response and how the phonons couple
to electrons close to Fermi energy.

Our calculated phonon density of states $F(\omega)$ and the Eliashberg
function $\alpha^{2}F(\omega)$ of bcc yttrium are shown in Fig. 6,
where we also show the corresponding values for hcp yttrium taken
from our earlier work \cite{pps_y1}. In the insets of Fig. 6, we
show the variations of $\omega$ and $\lambda_{\mathbf{q}\nu}$ of
bcc yttrium along $\Gamma$-N direction. A comparison of $F(\omega)$
of bcc yttrium with that of hcp yttrium clearly shows the hardening
of the bcc lattice, which has become almost three times as hard.  Some
of the significant additions to $F(\omega)$ of bcc yttrium take place
around the peak at $59$ meV, the peak itself arising out of the displacement
of yttrium atoms with components in all the three directions. We also
find that the phonon mode (with displacement having components along
$+y$ and $-z$ directions) in bcc yttrium along $\Gamma$-N direction
softens to 9 meV at $\mathbf{q}=(0.0,0.375,0.375)$ in units of $2\pi/a$,
and it has a value of 12.6 meV at the N symmetry point. Phonon softening
in fcc yttrium is discussed by Yin \emph{et al}. \cite{yin}. 

\begin{figure}
\begin{centering}\subfigure[]{\includegraphics[clip,scale=0.43]{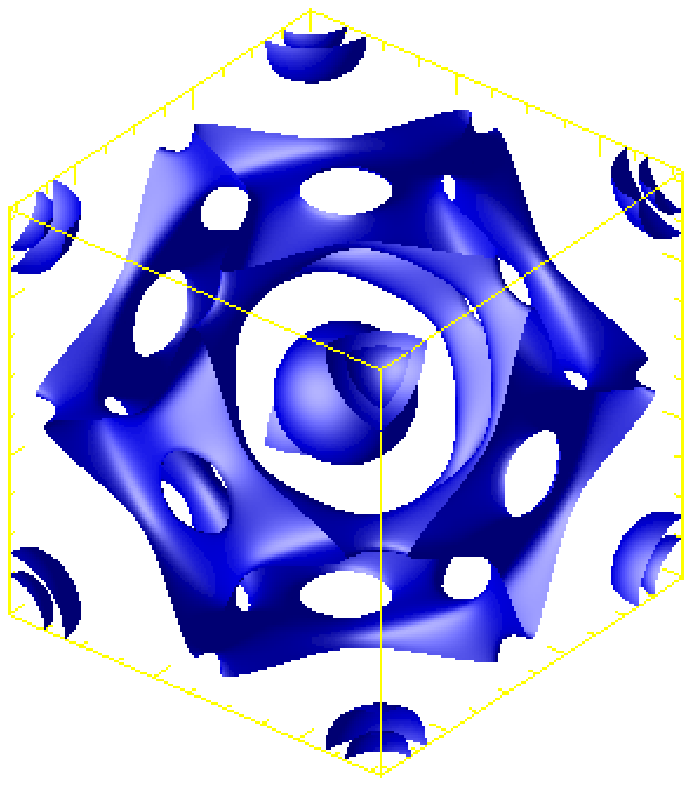}}
~~~\subfigure[]{\includegraphics[clip,scale=0.43]{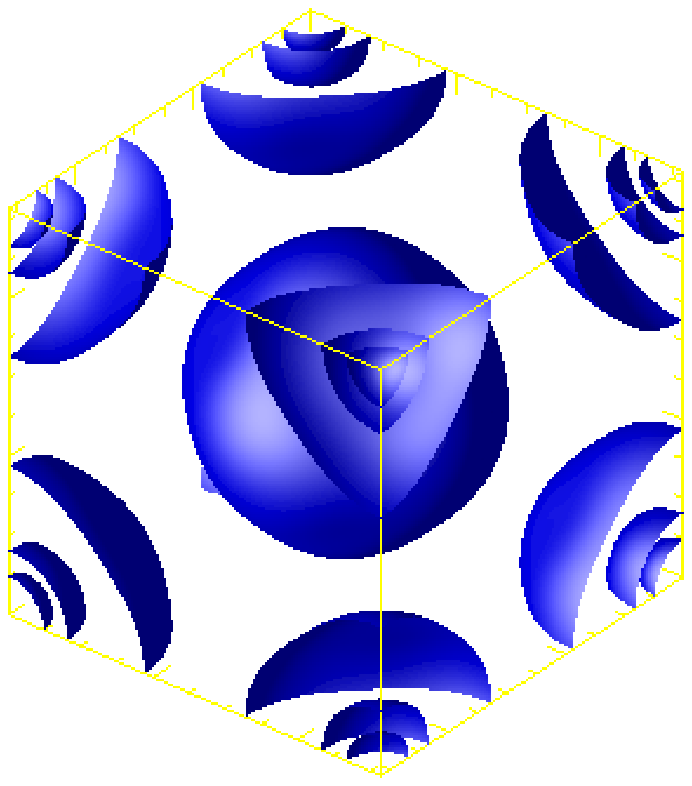}}\par\end{centering}

~

\begin{centering}\subfigure[]{\includegraphics[clip,scale=0.25,angle=-90]{fig5c.epsi}}~~~\subfigure[]{\includegraphics[clip,scale=0.25,angle=-90]{fig5d.epsi}}\par\end{centering}

\caption{The isosurfaces and contours of charge density within a 5 mRy energy
window around the Fermi energy in bcc yttrium. The values of the isosurfaces
are (a) 3.5 and (b) 4.5 in units of $10^{-2}[e/a.u.^{3}]$, which
is also the unit used in (c) and (d). }
\end{figure}

The pressure-induced increase in $\alpha^{2}F(\omega)$ in yttrium,
accompanied by structural phase transformations, seems to arise from
the enhancement of contributions from existing modes and coupling
strongly to the newly created modes. For example, the peaks around
$23$ and $27$ meV are created in the dhcp phase and they become
much stronger in the bcc phase, while the peak at $59$ meV is created
in the bcc phase. As a result, the average electron-phonon coupling
constant $\lambda$ increases from $0.46$ in hcp yttrium to $1.8$
in bcc yttrium.

Somewhat surprisingly, we find that the largest contribution of $\lambda_{\mathbf{q}\nu}\approx5$
to $\lambda_{\mathbf{q}}=5.9$ comes from the phonon mode along $\Gamma$-N
direction with $\mathbf{q}$ as mentioned above. At N symmetry point,
the same mode contributes $\lambda_{\mathbf{q}\nu}\approx2.9$ to
$\lambda_{\mathbf{q}}=5.2$. Smaller contributions with $\lambda_{\mathbf{q}\nu}>1$
also come from $\mathbf{q}$-points lying close to $\Gamma$$-$P
direction and to-wards H point. The fact that the smaller regions
in reciprocal space make substantial contributions to the electron-phonon
coupling, a precise determination of $\lambda$ using a uniform $\mathbf{q}$-grid
becomes computationally difficult. Since most of these contributions
come from the low energy phonons, it is not expected to have a significant
effect on the determination of $T_{c}$. To illustrate the point,
we also show in Fig. 6, the $F(\omega)$ and $\alpha^{2}F(\omega)$
of bcc yttrium calculated at the lattice constant $a=5.1987$ a.u.
as described earlier. Not surprisingly, the difference in $\lambda$,
resulting from the two sets of calculations, is more pronounced in
$15$-$25$ meV region. 

\begin{figure}
\begin{centering}\includegraphics[clip,scale=0.3]{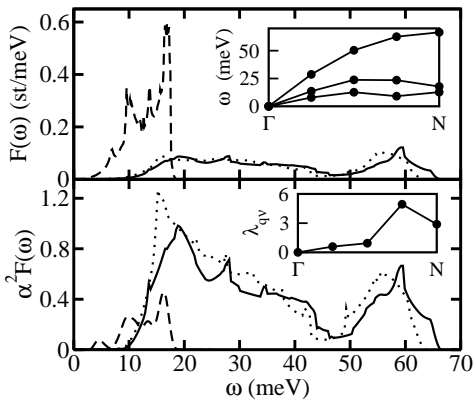}\par\end{centering}

\caption{The phonon density of states $F(\omega)$ (top panel) and the Eliashberg
function $\alpha^{2}F(\omega)$ (bottom panel) of bcc (solid line)
and hcp (dashed line) yttrium. The dotted line in the two panels show
$F(\omega)$ and $\alpha^{2}F(\omega)$ of bcc yttrium at $a=5.1987$
a.u., respectively. The two insets show $\omega$ and $\lambda_{q\nu}$(for
the lowest mode) along $\Gamma$-N direction. In the top panel, the
phonon density of states of hcp yttrium has been reduced by a factor
of 4 for clarity. }
\end{figure}

\begin{figure}
\begin{centering}\includegraphics[clip,scale=0.3]{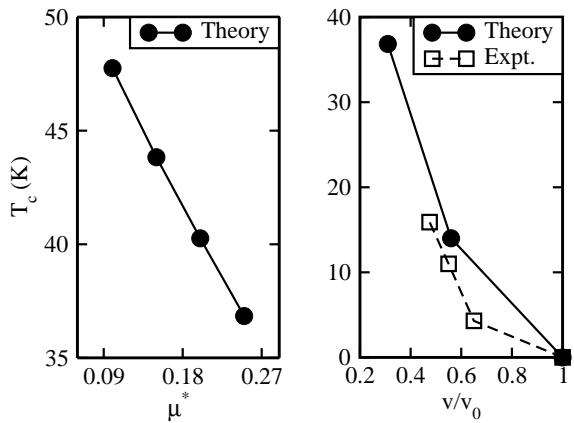}\par\end{centering}

\caption{The calculated $T_{c}$ (solid circles) as a function of $\mu^{*}$
(left panel) and $v/v_{0}$ (right panel). Experimentally observed
$T_{c}$ (open squares), taken from Ref. \cite{hamlin}, are also
shown in the right panel. The calculated $T_{c}$ for hcp and dhcp
phases of yttrium are taken from Ref. \cite{pps_y1}.}
\end{figure}

The most important result of the present work is obtained by solving
numerically the isotropic gap equation \cite{allen2,private1} using
the calculated Eliashberg function $\alpha^{2}F(\omega)$. The results
of such a calculation for bcc yttrium as a function of $\mu^{*}$
is shown in Fig. 7. We find that for $\mu^{*}$ ranging from $0.1$
to $0.25$, the $T_{c}$ values range from $47.7$ to $36.8$ K. A
comparison of our calculated $T_{c}$, including the ones described
in Ref. \cite{pps_y1}, with experiment is shown in Fig. 7. Thus,
the present work clearly shows the possibility of a $T_{c}$ of $30$
K or above for compressed, bcc yttrium. 

In conclusion, we have shown that compressed, bcc yttrium has a large
electron-phonon coupling with $\lambda=1.8$, leading to a superconducting
transition temperature $T_{c}\approx30$ K or above. We have also
shown that such a large coupling is a direct consequence of lattice
hardening and its subsequent coupling to Fermi electrons.

\end{document}